\documentclass{kapproc}
\usepackage{procps}
\usepackage[dvips]{graphicx}
\usepackage{amsmath}
\usepackage{graphicx}
\usepackage{amsfonts}
\usepackage{amssymb}
\normallatexbib

\begin{document}

\articletitle[Ultrafast real-time spectroscopy of low dimensional
CDW compounds]{Ultrafast real-time spectroscopy\\ of low
dimensional\\ charge density wave compounds}

\author{\underline{J. Demsar}$^1$, D. Mihailovic$^1$, V.V. Kabanov$^1$, and K. Biljakovic$^2$}
\affil{$^1$J. Stefan Institute, Jamova 39, 1000 Ljubljana,
Slovenia\\ $^2$Institute for Physics, Bijenicka 46, HR-10000
Zagreb, Croatia} \email{jure.demsar@ijs.si}

\begin{abstract}
We present a femtosecond time-resolved optical spectroscopy (TRS)
as an experimental tool to probe the changes in the low energy
electronic density of states as a result of short and long range
charge density wave order. In these experiments, a femtosecond
laser pump pulse excites electron-hole pairs via an interband
transition in the material. These hot carriers rapidly release
their energy via electron-electron and electron-phonon collisions
reaching states near the Fermi energy within 10-100 fs. The
presence of an energy gap in the quasiparticle excitation spectrum
inhibits the final relaxation step and photoexcited carriers
accumulate above the gap. The relaxation and recombination
processes of photoexcited quasiparticles are monitored by
measuring the time evolution of the resulting photoinduced
absorption. This way, the studies of carrier relaxation dynamics
give direct information of the temperature-dependent changes in
the low energy density of states. Here we present the application
of the femtosecond time-resolved optical spectroscopy for studying
changes in the low energy electronic density of states in low
dimensional charge density wave systems associated with various
charge density wave (CDW) transitions and review some recent
experiments on quasi 1D and 2D CDW compounds.
\end{abstract}


\section{Introduction}

\begin{figure}[h]
\begin{center}
\includegraphics[width=11cm
]%
{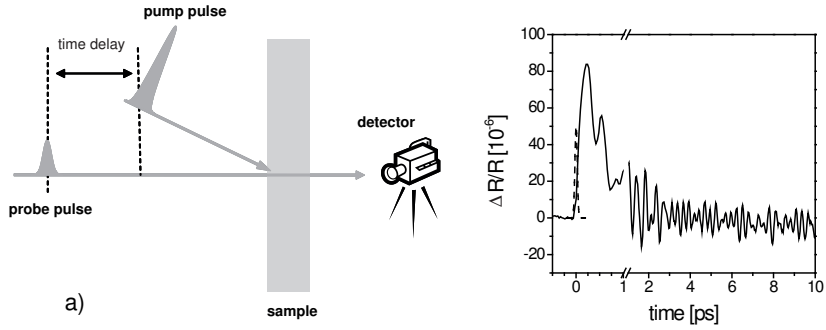}%
\caption{a) Schematic of the experimental technique: pump pulse
excites the system under investigation, while the resulting
changes in optical properties (reflectivity, transmission) as a
function of time are investigated by a suitably delayed probe
pulse. Panel b) presents the photoinduced reflectivity change in
quasi-2D charge density wave compound 1T-TaS$_{2}$ obtained at 10
K.
The dashed line represents the excitation pulse.}%
\label{Setup}%
\end{center}
\end{figure}

Femtosecond time-resolved optical spectroscopy has been shown in
the last couple of years to present an excellent alternative to
the more conventional time-averaging frequency-domain
spectroscopies for probing the changes in the low energy
electronic structure in strongly correlated systems
\cite{Kabanov,ACS}. In these experiments (see Figure \ref{Setup}),
a femtosecond laser pump pulse excites electron-hole pairs via an
interband transition in the material. In a process which is
similar in most materials including metals, and superconductors,
these hot carriers rapidly release their energy via
electron-electron and electron-phonon collisions reaching states
near the Fermi energy within 10-100 fs. Further relaxation and
recombination dynamics, determined by measuring photoinduced
changes in optical properties (reflectivity, transmissivity or, in
case of time-resolved terahertz spectroscopy, far infrared
conductivity) as a function of time after photoexcitation, depends
strongly on the nature of the low-lying electronic spectrum. In
particular, the experimental technique was found to be sensitive
to opening of the superconducting gap, appearance of a short-range
and long range charge-density wave order \cite{TRbb,TR2DCDW}, and
changes in the electronic specific heat and electron-phonon
coupling associated with the heavy fermion behavior \cite{HF},
just to mention a few. What is particularly important is the fact,
that even though the probe photon wavelength in these experiments
ranges from THz \cite{Averitt,Averitt2,JDMgB2} (enabling direct
measurement of photoinduced conductivity dynamics), mid-IR to
several eV [1,9-16], the dynamics is in many instances the same
\cite{Averitt,Averitt2,JDMgB2}, supporting the idea \cite{Kabanov}
that the photoinduced reflectivity (transmissivity) dynamics is
determined by relaxation and recombination processes of
quasiparticles in the vicinity of Fermi energy.

Since the optical penetration depth in these materials is on the
order of 100 nm, the technique is essentially a bulk probe.
Moreover, since the effective shutter speed is on the order of a
picosecond, the technique is particularly useful to probe the
systems with (dynamic) spatial inhomogeneities. In this case
different local environments (that appear frozen on the timescale
of picoseconds) give rise to different components in measured
photoinduced reflectivity (transmissivity) traces. As the
different components can have different time
scales\cite{Y124,ODpaper}, temperature\cite{Kabanov,TRbb},
photoexcitation intensity, and probe polarization\cite{Y124} or
wavelength dependences, they can be easily extracted.

Furthermore, due to the fast effective shutter speed of the
technique ($\approx1$ ps) one can expect to observe short lived
fluctuations that would appear frozen on this timescale. Indeed,
the experiments on quasi-1D charge density wave compound
K$_{0.3}$MoO$_{3}$ \cite{TRbb} suggest that above the transition
temperature to the 3D ordered CDW state the technique is sensitive
to the presence of short range 3D fluctuations of CDW order.

\section{Experimental details}

In the experiments discussed below, a Ti:sapphire mode-locked laser operating
at a 78 MHz\ repetition rate and pulse length of 50-70 fs was used as a source
of both pump and probe pulse trains. The wavelength of the pulses was centered
at approximately $\lambda$ $\approx$ 800nm (1.58eV) and the intensity ratio of
pump and probe pulses was about 100:1. The pump and probe beams were crossed
on the sample's surface, where the angle of incidence of both beams was less
than 10$^{o}$. The diameters of \ the beams on the surface were\ $\sim$%
100$\mu$m for the pump beam and $\sim$50$\mu$m for the probe beam. The typical
energy density of pump pulses was $\sim0.1-1$ $\mu$J$/$cm$^{2}$, which
produced a weak perturbation of the electronic system with the density of
thermalized photoexcited carriers on the order of $10^{-4}-10^{-3}$ per unit
cell (the approximation is based on the assumption that each photon with
energy $\hbar\omega$ creates $\hbar\omega/\Delta$ thermalized photoexcited
carriers, where $\Delta\approx50$ meV is of the order of the CDW gap
\cite{TRbb}). The train of the pump pulses was modulated at 200kHz \ with an
acousto-optic modulator and the small photoinduced changes in reflectivity or
transmission were resolved out of the noise with the aid of phase-sensitive
detection. The pump and probe beams were cross-polarized to reduce scattering
of pump beam into the detector (avalanche photodiode).\ This way, photoinduced
changes in reflectance of the order of $10^{-6}$ can be resolved. A detailed
description of the experimental technique can be found in Ref. \cite{ACS}.

\section{Photoexcited quasiparticle dynamics in narrow-gap materials}

In this section we review the basic ideas of the theoretical model
\cite{Kabanov} adopted to associate the measured amplitude and the relaxation
dynamics of the photoinduced transients with the corresponding changes in the
low-energy electronic structure in these narrow-gap materials.%

\begin{figure}[h]
\begin{center}
\includegraphics[
width=9.5cm
]%
{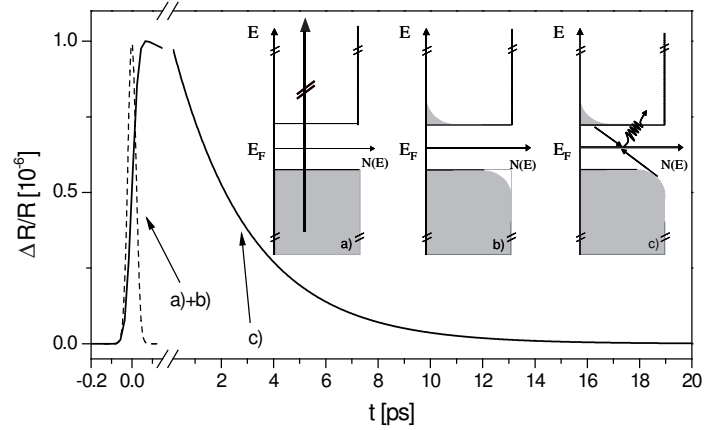}%
\caption{Schematic of the photoexcitation, carrier relaxation and
photoinduced absorption processes in materials with a small energy
gap in DOS. Interband photoexcitation (step a) is followed by an
ultrafast initial e-e thermalization (step b). The gap in the
density of states creates a relaxation bottleneck. The probe pulse
measures through the photoinduced absorption (step
c) the time evolution of the photoexcited carrier density.}%
\label{Processes}%
\end{center}
\end{figure}

The basis of the model is shown in Figure \ref{Processes}. A pump pulse with
photon energy $\hbar\omega$ (e.g. 1.5 eV) excites carriers from occupied
states below $E_{F}$ to unoccupied states in bands $\hbar\omega$ above -
schematically shown by \textbf{a)}. The initial phase of the
photoexcited\ carrier relaxation after absorption of the pump laser photon
proceeds rapidly. The photoexcitation is followed by carrier thermalization
through electron-electron scattering with a characteristic time $\tau
_{e-e}\sim\frac{\hbar E_{F}}{2\pi E^{2}}$ for intraband relaxation, where $E$
is the carrier energy measured from\ the Fermi energy $E_{F}$. In the initial
thermalization process, a quasiparticle avalanche multiplication due to
electron-electron collisions takes place as long as $\tau_{e-e}$ is less than
the electron-phonon (e-ph) relaxation time $\tau_{e-ph}$ which is $\approx100$
fs in these materials. Therefore, in the absence of gap in the density of
states the photoexcitation is followed by rapid carrier relaxation resulting
in slightly elevated electron-phonon temperature within $\sim100$ fs.

When the gap with magnitude $2\Delta$ is present in the low energy density of
states, phonons with energies less than $2\Delta$ cannot contribute to the
relaxation of carriers just above the gap, therefore the situation is strongly
modified and a bottleneck in the relaxation occurs after $t\sim100$ fs. As a
result quasiparticles accumulate near the gap, forming a non-equilibrium
distribution shown by \textbf{b)}. Since typical values of the energy gap in
CDW's (and cuprates) are of the order of $2\Delta\sim$ 30 - 100 meV, each
photon thus creates $20\sim40$ quasiparticles given by $\hbar\omega/2\Delta
$.\ The final relaxation step across the gap is strongly suppressed
\cite{RothwarfTaylor}, since the high energy phonons emitted by bi-particle
recombination (shown by \textbf{c)}) have enough energy to further excite
electron-hole pairs. Therefore the quasiparticles together with\emph{\ }high
frequency phonons (with $\hbar\omega_{q}>2\Delta$) form a near-steady state
distribution. The recovery dynamics of this system is governed by the decay
$\hbar\omega_{q}>2\Delta$ phonon population, governed either by the diffusion
out of the probed volume or by anharmonic decay to $\hbar\omega_{q}<2\Delta$
phonons. Since in CDW compounds the energy gap is on the order of 30 - 100
meV, the phonons in question are optical phonons, whose anharmonic decay times
are on the order of picoseconds the later mechanism was found to be dominant
(diffusion takes place on the timescale of 100's of picoseconds).

Considering the probe process, we are measuring photoinduced changes of
transmission or reflectivity on the picosecond timescale. Since we are dealing
with weak perturbations, we can assume that the photoinduced transmission
$\Delta T/T$ (or reflectivity $\Delta R/R$) is in the linear approximation
proportional to photoinduced absorption $\Delta A/A$. Through the Fermi golden
rule the photoinduced absorption is due to changes in the \emph{initial or
final state} carrier density. As the lifetime of quasiparticles high above
$E_{F}$ is of the order of 10 fs, we can assume that the main changes in
absorption involve photoexcited carriers just above the gap as initial or
final states for absorption. Since the probe laser photon energy $\hbar\omega$
is typically well above the plasma frequency, we make an approximation that
the photoinduced absorption is given by the Fermi golden rule, with
photoinduced quasiparticles above the gap as initial states and unoccupied
states at $\hbar\omega$ above the Fermi energy as final states - see
\textbf{c)} in insert to Figure \ref{Processes}. The amplitude of the
photoinduced absorption is thus proportional to the photoexcited quasiparticle
density $n_{pe}$ and by measuring the photoinduced transmission $\Delta T/T$
(or reflectivity $\Delta R/R$) the temporal evolution of the photoexcited
carrier density $n_{pe}$ is probed. The photoinduced transmission amplitude is
weighted by the dipole matrix element and the joint density of states, so
$\Delta T/T\propto-n_{pe}\rho_{f}\left\vert M_{ij}\right\vert ^{2}$, where
$n_{pe}$ is the photoexcited carrier density, $\rho_{f}$ is the density of
(final) unoccupied states, and $M_{ij}$ is the dipole matrix element.

Based on the above arguments, that after initial e-e and e-ph thermalization
processes quasiparticles and $\hbar\omega_{q}>2\Delta$ phonons are is
quasi-thermal equilibrium, and that the amplitude of the transient is
proportional to the photoinduced quasiparticle density, the one-to-one
relation between the temperature dependence of the photoinduced transient
amplitude and the amplitude of the low-energy gap in the DOS can be found.
Assuming that the energy gap is isotropic, one can approximate non-equilibrium
phonon $(n_{\omega_{q}})$ and quasiparticle $(f_{\varepsilon})$ distribution
functions as follows \cite{Kabanov,Aronov}:
\begin{equation}
n_{\omega_{q}}=\Biggm\{
\genfrac{}{}{0pt}{}{\frac{1}{\exp(\frac{\hbar\omega_{q}}{k_{B}T})-1}%
\quad,\quad\hbar\omega_{q}<2\Delta}{\frac{1}{\exp(\frac{\hbar\omega_{q}}%
{k_{B}T^{\prime}})-1}\quad,\quad\hbar\omega_{q}>2\Delta}
\label{Nph}%
\end{equation}%
\begin{equation}
f_{\varepsilon}=\quad\frac{1}{\exp(\frac{\varepsilon}{k_{B}T^{\prime}})+1}\;,
\end{equation}
where $T$ is the lattice temperature and $T^{\prime}$ is the temperature of
quasiparticles and high frequency phonons with $\hbar\omega_{q}>2\Delta$. The
number of photoexcited quasiparticles $n_{pe}$ can be calculated as the
difference between the numbers of thermally excited quasiparticles (per unit
cell) after and before photoexcitation characterized by temperatures
$T^{\prime}$ and $T$. The number of photoexcited carriers $n_{pe}%
(=n_{T^{\prime}}-$ $n_{T})$ can be obtained directly considering energy
conservation \cite{Kabanov}.

As an illustration of the calculation, let us assume that $\Delta=\Delta^{p}$
is temperature \emph{independent} and large in comparison to $k_{B}T$. Since
the magnitude of the gap $\Delta^{p}$ is of the order of several 10 meV, which
corresponds to temperatures of a few hundred kelvins, we can assume that
quasiparticles are non-degenerate and $f_{\varepsilon}$ can be approximated as
$f_{\varepsilon}$ $\sim\exp\left(  -\varepsilon/k_{B}T\right)  $. Similarly we
can approximate $n_{\omega_{q}}\sim\exp(-\hbar\omega_{q}/k_{B}T)$. When
considering the temperature-independent (pseudo-)gap, we take the
quasiparticle density of states given by
\begin{equation}
N\left(  E\right)  =\Biggm\{%
\genfrac{}{}{0pt}{}{0\;,\;E<\Delta^{p}}{N\left(  0\right)  \;,\;E>\Delta^{p}}%
.
\end{equation}
Strictly speaking the model density of states corresponds to the real gap,
however $\Delta^{p}$ could be understood also as an energy where the
relaxation of photoexcited quasiparticles in inhibited. E.g., the density of
states below $\Delta^{p}$ could be finite but the relaxation through these
states is suppressed (e.g. relaxation through localized states) therefore they
are not available for the relaxation.

Further, we assume that the phonon spectral density is constant at large
frequencies $(\hbar\omega_{q}>2\Delta^{p})$. In this case the quasiparticle
energy and the energy of high frequency phonons at temperature T are given by
\[
E_{T}=2N\left(  0\right)  \Delta^{p}T\exp(-\Delta^{p}/k_{B}T)\text{ \ ;
\ \ }E_{T}^{ph}=\frac{2\nu\Delta^{p}T}{\Omega_{c}}\exp(-2\Delta^{p}/k_{B}T)
\]
respectively. Here $\nu$ is the number of high frequency phonon modes (per
unit cell) and $\Omega_{c}$ is the phonon cut-off frequency. Since we assume
that after photoexcitation the high energy phonons and quasiparticles are
described by the same temperature ($T^{\prime}$), we write the conservation of
energy as%

\begin{equation}
\mathcal{E}_{I}\;=\left(  E_{T^{\prime}}+E_{T^{\prime}}^{ph}\right)  -\left(
E_{T}+E_{T}^{ph}\right)  \label{38}%
\end{equation}
where $\mathcal{E}_{I}$ is the energy density per unit cell deposited by the
incident pump laser pulse. Since quasiparticle density is given by
$n_{T}=2N\left(  0\right)  T\exp(-\Delta^{p}/k_{B}T)$ \cite{Aronov}, by making
the approximation that $k_{B}T\ll\Delta$, Eq.(\ref{38}) can be rewritten in
terms of quasiparticle densities at temperatures $T^{\prime}$ and $T$%

\begin{equation}
\left(  n_{T^{\prime}}-n_{T}\right)  \Delta^{p}+\left(  n_{T^{\prime}}%
^{2}-n_{T}^{2}\right)  \frac{\nu\Delta^{p}}{2\hbar\Omega_{c}N(0)^{2}%
k_{B}T^{\prime}}=\mathcal{E}_{I}\;. \label{energy}%
\end{equation}
There are two limiting cases to be considered with respect to the ratio of
photoexcited vs. thermally excited quasiparticle densities, $n_{pe}/n_{T}$.

In the low temperature limit $n_{pe}\gg n_{T}$, since $n_{T}$ is exponentially
small. In this case $n_{T^{\prime}}\gg n_{T}$ and by equaling $n_{T}$ to 0 in
Eq.(\ref{energy}) one gets the quadratic equation for $n_{T^{\prime}}$
($\simeq n_{pe}$). Since $n_{pe}$ is small, one can neglect the quadratic term
in Eq.(\ref{energy}) obtaining
\begin{equation}
n_{pe}\left(  =n_{T^{\prime}}-n_{T}\right)  =\mathcal{E}_{I}/\Delta
^{p}\mathbf{\;.} \label{lowTNph}%
\end{equation}
It follows that in the low temperature limit the photoinduced signal amplitude
$\left(  \approx n_{pe}\right)  $ is independent of temperature and its
magnitude is proportional to photoexcitation intensity $\mathcal{E}_{I}$.

The second limiting situation is the case when $n_{pe}\ll n_{T}$ (high
temperature limit). Then, taking into account that $n_{pe}=(n_{T^{\prime}%
}-n_{T})\ll n_{T}$ and $n_{T}=2N(0)k_{B}T\exp(-\Delta^{p}/k_{B}T)$, the number
of photogenerated quasiparticles at temperature $T$ is given by%

\begin{equation}
n_{pe}=\frac{\mathcal{E}_{I}/\Delta^{p}}{1+\frac{2\nu}{N(0)\hbar\Omega_{c}%
}\exp(-\Delta^{p}/k_{B}T)}\ . \label{NphBE}%
\end{equation}
It is important to stress that Eq.(\ref{NphBE}) includes also the solution of
Eq.(\ref{energy}) in the low temperature limit given by Eq.(\ref{lowTNph}).

Similar derivation can be applied to determine the T-dependence of the number
of photoexcited carriers in the case of a temperature dependent
mean--field-like gap $\Delta_{c}(T)$ such that $\Delta_{c}(T)$ $\rightarrow0$
as $T\rightarrow T_{c}$. This results in a slightly modified expression for
$n_{pe}$ that again contains both the low and the high temperature limits%

\begin{equation}
n_{pe}=\frac{\mathcal{E}_{I}/(\Delta_{c}(T)+k_{B}T/2)}{1+\frac{2\nu}%
{N(0)\hbar\Omega_{c}}\sqrt{\frac{2k_{B}T}{\pi\Delta_{c}(T)}}\exp(-\Delta
_{c}(T)/k_{B}T)}. \label{NphBCS}%
\end{equation}

Note that in Eqs.(\ref{NphBE}) and (\ref{NphBCS}) the explicit form of
$n_{pe}\left(  T\right)  \,$depends only on the ratio $k_{B}T/\Delta\left(
0\right)  $, showing that the intensity of the photoresponse is a universal
function of $k_{B}T/\Delta$ as long as the particular functional form of
temperature dependence of $\Delta$ is the same. The only parameter in
Eqs.(\ref{NphBE}) and (\ref{NphBCS}) is the dimensionless constant $\frac
{2\nu}{N(0)\hbar\Omega_{c}}$, which can be estimated for each compound studied
to at least an order of magnitude. Therefore by fitting the T-dependence of
the PI amplitude of the transient the value of the gap $\Delta$ can be
determined quite accurately.%

\begin{figure}
[h]
\begin{center}
\includegraphics[width=4.4in
]%
{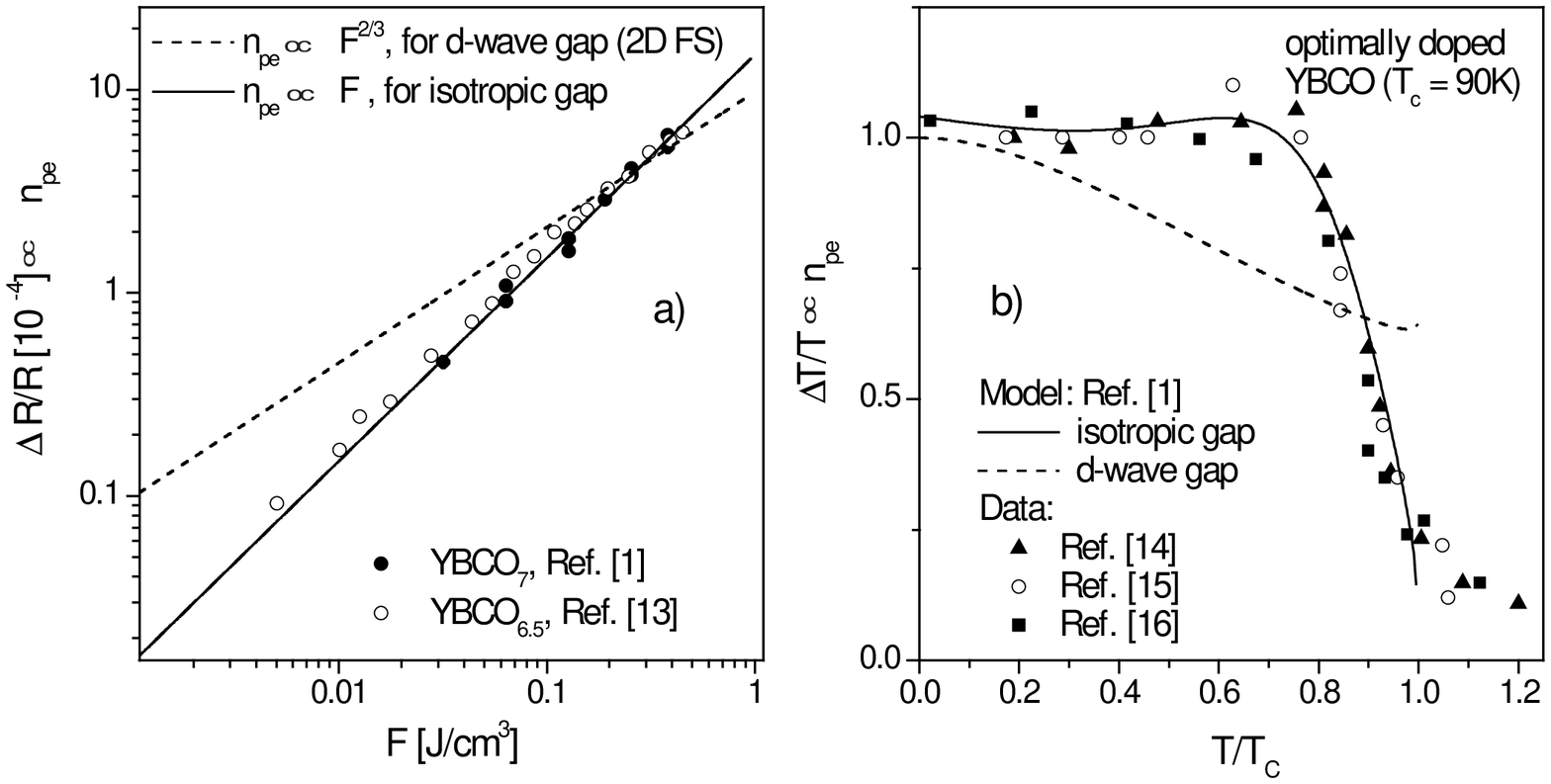}%
\caption{a) Photoexcitation intensity (F) dependence of the
photoinduced transient amplitude from Refs. \cite{Kabanov,Segre}.
The data, presented on log-log plot show linear dependence over
two orders of magnitude in F (linear fit is given by solid line).
On the other hand, the expected intensity dependence for
anisotropic gap in 2D gives F$^{2/3}$ dependence \cite{Kabanov}
(best fit shown by dashed line). b) the experimental T-dependence
of photoinduced transient amplitude from \cite{Stevens,Han,EPL}
compared to the prediction for the case of large isotropic gap
(solid line), and anisotropic gap with nodes (dashed), given by
Eqs. 6 and 12 from Ref. \cite{Kabanov}
respectively.}%
\label{dwaveFig}%
\end{center}
\end{figure}

The above derivation has been extended also for the case of an anisotropic gap
with nodes \cite{Kabanov}, which could be used to describe the quasiparticle
relaxation dynamics in cuprates, where the vast amount of data suggests d-wave
order parameter with nodes\cite{Tsuei}. However, the experimental results on
cuprates were found to be at odds with the simple d-wave picture. In
particular, linear intensity dependence of the photoinduced transient
amplitude and its peculiar temperature dependence - see Figure \ref{dwaveFig}
are inconsistent with simple d-wave case scenario. The fact that these
experiments do suggest large (more or less isotropic) gap in the density of
states, can be however due to particularities of dynamics in cuprates. In
other words, since time-resolved techniques measure the fastest channel for
relaxation, it is possible that carrier relaxation from antinodes (in
direction of maximum of the gap) to nodes is much slower process than
recombination to the condensate.

To complete the description of the theoretical model for carrier relaxation
dynamics in narrow-gap systems, we should briefly discuss the relaxation
dynamics. We have suggested that the lifetime of photoexcited quasiparticle
density is governed by anharmonic decay of high frequency phonons. Using the
kinetic equations for phonons taking into account phonon-phonon scattering
\cite{Lifshitz}we have obtained the following expression for the rate of
recovery of the photoexcited state \cite{Kabanov}.
\begin{equation}
\tau^{-1}=\frac{12\Gamma_{\omega}k_{B}T^{\prime}\Delta(T)}{\hbar\omega^{2}}\;.
\label{e25}%
\end{equation}

The relaxation time for the temperature dependent gap $\Delta_{c}\left(
T\right)  $ is expected to show a divergence $\tau\propto1/\Delta_{c}(T)$
$\rightarrow\infty$ due to the gap closing as T$_{c}$ is approached from
below. This can be easily understood considering phase space arguments.
Namely, in the case of a mean-field-like gap, upon increasing temperature
closer to $T_{c}$, the specific heat of high frequency phonons increases while
the specific heat of $\hbar\omega_{q}<2\Delta$ phonons is decreasing, giving
rise to $\propto1/\Delta_{c}(T)$ increase in relaxation time \cite{Kabanov}.

\section{Carrier relaxation dynamics in quasi-1D CDW compounds.}

In these section we briefly review recent experimental results on femtosecond
time-resolved spectroscopy on CDW compounds \cite{TRbb,TR2DCDW}. In
particular, we focus on the results on\ quasi-1D CDW compound K$_{0.3}%
$MoO$_{3}$, while for the details on the studies of quasi-2D CDW
are given elsewhere \cite{TR2DCDW}.

\begin{figure}[h]
\begin{center}
\includegraphics[width=4.1in
]%
{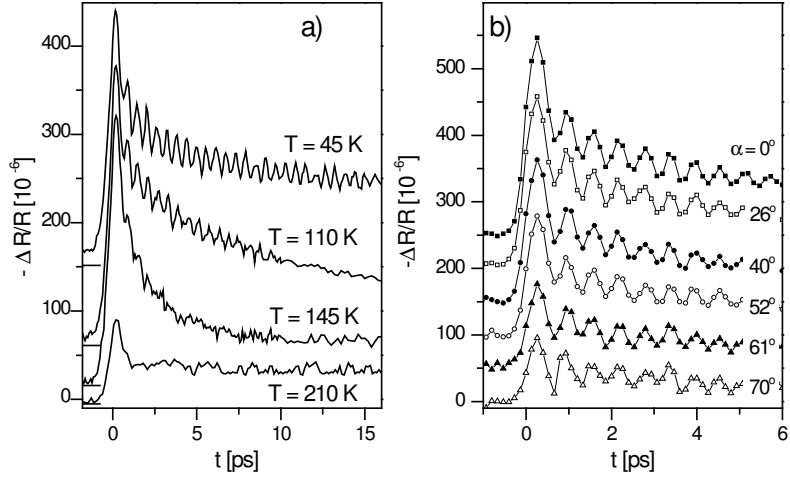}%
\caption{ a) Photoinduced reflectivity traces taken on
K$_{0.3}$MoO$_{3}$ at temperatures below and above
T$_{c}^{3D}=183$ K. b) The dependence of the photoinduced signal
on the probe polarization, with $\alpha$ being the angle between
the polarization of probe beam and the crystal [102] axis. The
data
was taken at 100 K. }%
\label{BBRaw}%
\end{center}
\end{figure}

Molybdenum oxides A$_{0.3}$MoO$_{3}$, where A is a monovalent metal like K,
Rb, or Tl - also called blue bronzes due to their shiny blue appearance - are
well known for their interesting electronic properties arising from their
one-dimensional (1D)\ chain structures \cite{GrunerReview}. K$_{0.3}$MoO$_{3}$
crystallizes in a monoclinic unit cell \cite{Xray}. The structure contains
rigid units comprised of clusters of ten distorted MoO$_{6}$ octahedra,
sharing corners along monoclinic $b$-axis. This corner sharing provides an
easy path for the conduction electrons along the chain direction. The chains
also share corners along the [102] direction and form infinite slabs separated
by the potassium cations. The [102] direction together with [010] direction
form a cleavage plane. At room temperature K$_{0.3}$MoO$_{3}$ is a highly
anisotropic one-dimensional metal with conductivity ratios $\sigma_{b}%
:\sigma_{2a-c}:\sigma_{2a+c}=30:1:0.05$ (it is a quasi 1D
metal).\cite{GrunerReview} Upon cooling, blue bronzes become
susceptible to a Peierls instability on the 1D chains causing
fluctuating local CDW ordering. Upon further cooling, inter-chain
interactions cause the CDWs on individual chains to become
correlated, eventually undergoing a second-order phase transition
to a three-dimensionally (3D) ordered state below $T_{c}^{3D}=183$
K. The formation of a 3D\ CDW ordered state is concurrent with the
appearance of a gap $\Delta_{CDW}$ in the quasiparticle excitation
spectrum, while the collective excitations of the 3D CDW state are
described by an amplitude mode (AM) and a phase mode (phason)
\cite{GrunerBook}.

\begin{figure}
[h]
\begin{center}
\includegraphics[width=10cm
]%
{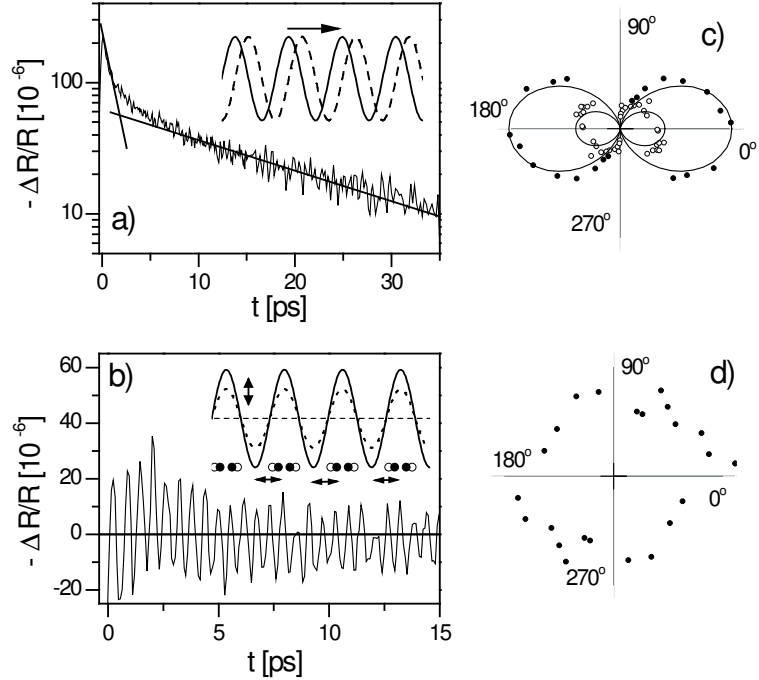}%
\caption{Below T$_{c}^{3D}$ the response consists of three
distinct components: sub-picosecond and 10 ps timescale dynamics
shown on a semi-log plot in a) and b) the oscillatory transient
with frequency 1.7 THz. While the sub-picosecond response is
attributed to the PI absorption on quasiparticles, the oscillatory
component and the 10 ps transient are due to photoexcitation of
collective modes (amplitudon and phason). The collective modes (q
= 0) are schematically shown in insets to panels a) and b)), where
solid lines and dots represent the unperturbed carrier density and
the ionic positions while dashed lines (open symbols) represent
the effect of photoexcitation. c) The amplitude of the fast
transients as a function of probe polarization with respect to the
crystal [102] direction below (solid ) and above (open circles)
$T_{c}^{3D}=183$ K. d) c) The amplitude of the oscillatory signal
as a function of probe polarization with respect to the
crystal [102] direction. }%
\label{BBkomponents}%
\end{center}
\end{figure}

Figure \ref{BBRaw}a) presents photoinduced reflectivity $\Delta
R/R$ as a function of time at different temperatures. Below
$T_{c}^{3D}$, an oscillatory component is observed on top of a
negative induced reflection, the latter exhibiting a fast initial
decay followed by a slower decay - see Figure \ref{BBkomponents}
a). As $T_{c}^{3D}$ is approached from below, the oscillatory
signal disappears, while the fast transient signal remains
observed well above $T_{c}^{3D},$ as shown by the trace at 210K.
For a quantitative analysis, we separate the different components
of the signal according to their temperature dependence and probe
polarization anisotropy.

Figure \ref{BBkomponents} presents the decomposed reflectivity
transient taken at T $=45$ K. Panel a) presents the signal with
the oscillatory component subtracted. The logarithmic plot enables
us to clearly identify two components with substantially different
lifetimes, one with $\tau_{s}\simeq$ 0.5 ps, and the other with
$\tau_{p}\gtrsim10$ ps at low $T.$ Their amplitudes and relaxation
times are analyzed by fitting the recovery dynamics with two
exponential decay. Since the two components have very different
temperature dependencies they are attributed to the quasiparticle
recombination (sub-picosecond component) and an overdamped
\emph{phason} relaxation (10 ps component),
respectively\cite{TRbb}. Importantly, the ps transient has a
pronounced probe polarization anisotropy with respect to the
crystal axes - plotted in panel c). Panel b) presents the
oscillatory component, whose Fourier spectrum shows a peak at
$\nu_{A}=$ 1.7 THz. In contrast to the transient signal, the
amplitude of the oscillatory signal is independent of polarization
- see panel d).

\subsection{Collective modes response}

\begin{figure}[h]
\begin{center}
\includegraphics[width=12cm
]%
{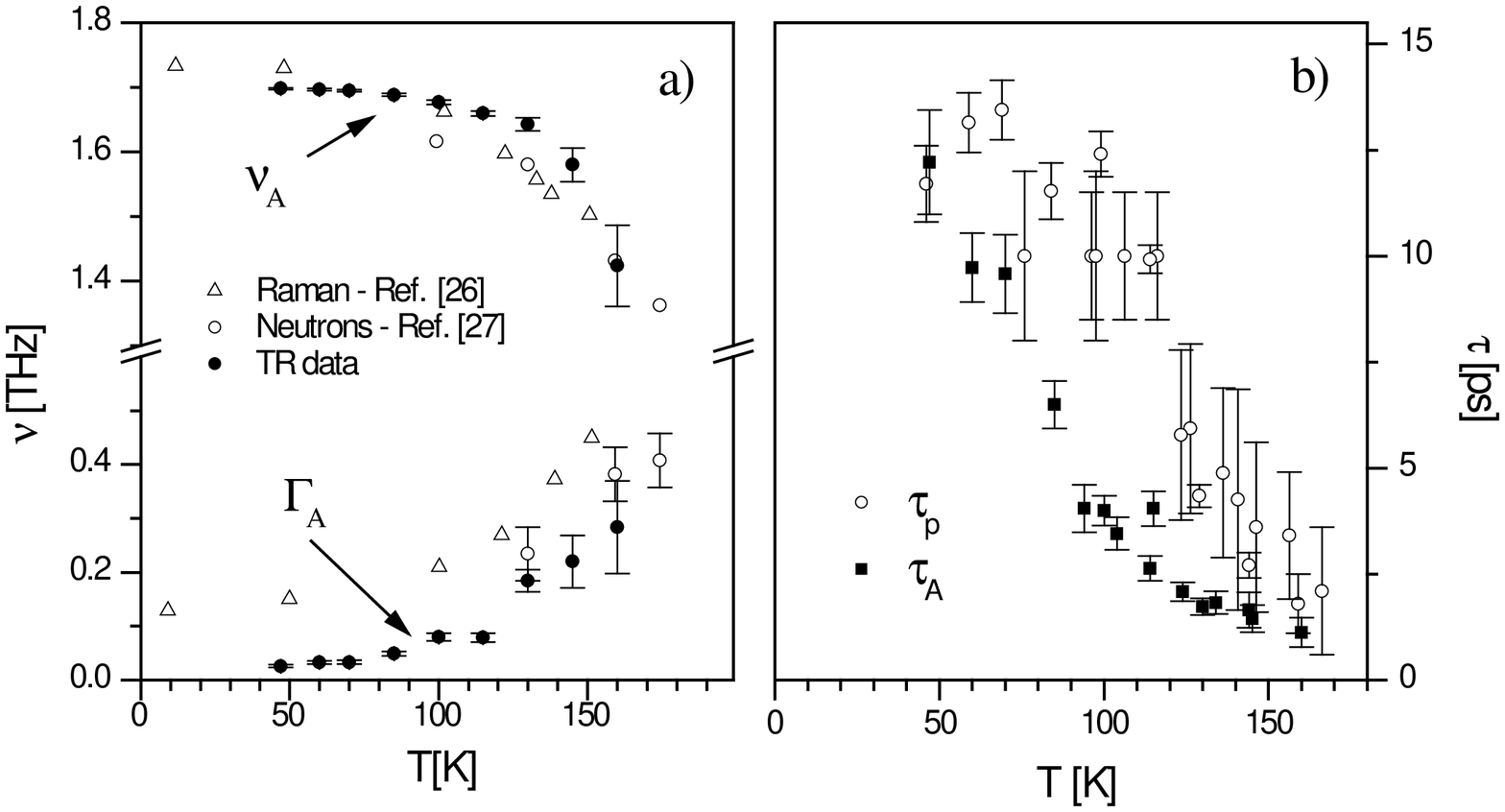}%
\vspace{-1.5cm} \caption{a)\ The oscillation frequency $\nu_{A}$
(full circles) and damping constant $\Gamma_{A}=1/(\pi\tau_{A})$
as functions of temperature (full circles). The data from Refs.
\cite{PougetNeutron} (open circles) and \cite{WachterRaman} (open
triangles) are also included for comparison. b) The phason damping
constant $\tau_{p}$ as a function of temperature (open circles),
compared to the amplitudon decay time $\tau_{A}$ is also plotted
for
comparison (solid squares).}%
\label{CollModes}%
\end{center}
\end{figure}

The oscillatory component, that has been observed also in the experiments on
the quasi-2D CDW compounds \cite{TR2DCDW}, has a pronounced $T$-dependence.
The frequency $\nu_{A}$ shows clear softening as T$_{c}$ is approached from
below, while damping $\Gamma_{A}=1/(\pi\tau_{A})$ increases - see Fig.
\ref{CollModes}. The measured $\nu_{A}$ and $\Gamma_{A}$ closely follow the
expected behavior for the \emph{amplitude mode} and are in good agreement with
previous spectroscopic neutron \cite{PougetNeutron} and Raman data
\cite{WachterRaman}.

The amplitudon is of $A_{1}$ symmetry and involves displacements
of ions about their equilibrium positions $Q_{0}$, which depend on
the instantaneous surrounding electronic density $n(t)$. Since the
excitation pulse is shorter than $h/\nu_{A},$ the excitation (and
subsequent ultrafast e-e thermalization) may be thought of as a
$\delta$-function-like perturbation of the charge density and the
injection pulse acts as a time-dependent displacive excitation of
the ionic equilibrium position $Q_{0}(t).$ The response of the
amplitudon to this perturbation is a modulation of the
reflectivity\ $\Delta R_{A}/R$ of the form $A(T)e^{-t/\tau
_{A}}\cos(\omega_{A}t+\phi_{0})$ by the displacive excitation of
coherent phonons (DECP) mechanism, known from femtosecond
experiments on semiconductors
\cite{Zeiger}.%

\vspace{-1.5cm}

While the association of the oscillatory component to the photoexcited
amplitude mode is straight-forward, the association of the 10 ps transient
with the overdamped phase mode needs further clarification. In equilibrium,
the phason mode is expected to be pinned \cite{GrunerBook}\ and at a finite
frequency $\omega_{p}>0$ \cite{KimBB}, but in non-equilibrium situation such
as here, where the excess carrier kinetic energy may easily exceed the
de-pinning energy, the mode may be de-pinned. In this case we may expect an
\textit{overdamped} reflectivity transient that can be written as
$\mathcal{P}(T)e^{-t/\tau_{p}}\cos(\omega_{p}t+\phi),$ with $\omega
_{p}\rightarrow0$, but with the damping constant which is expected to be
similar to that of the amplitude mode $\tau_{p}\simeq\tau_{A}$, i.e. $\sim$10
ps \cite{Tutis}. In Fig. \ref{CollModes} b) we plot $T$-dependence of
$\tau_{p}$. At $T=$50 K $\tau_{p}=12\pm2$ ps in agreement with the
$\Gamma=0.05\sim0.1$ THz linewidths of the pinned phason mode in microwave and
IR experiments \cite{GrunerReview,PougetNeutron,Degiorgi}. Indeed, comparing
$\tau_{p}$ and $\tau_{A}$ [also plotted in Fig. \ref{CollModes} b)], at 50K
$\tau_{p}\simeq\tau_{A}$ but the fall-off at higher temperatures appears to be
faster for $\tau_{A}$ than for $\tau_{p}.$With increasing temperature
$\tau_{p}$ is approximately constant up to 100 K and then falls rapidly as
$T\rightarrow T_{c}^{3D}$. The decrease of $\tau_{p}$ near $T_{c}^{3D}$ is
consistent with increasing damping due to the thermal\ phase fluctuations
arising from coupling with the lattice and quasiparticle excitations.

\subsection{Quasiparticle response}

\begin{figure}
[h]
\begin{center}
\includegraphics[width=4.5455in
]%
{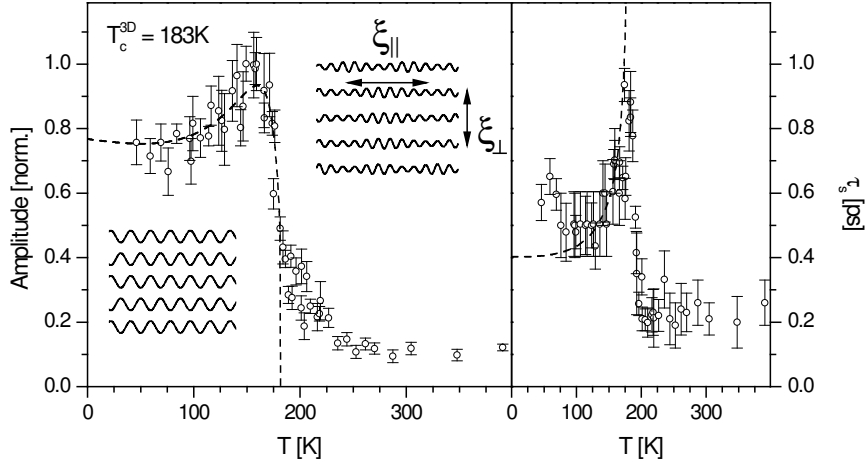}%
\caption{The temperature dependence of a) the amplitude and b) the relaxation
time of the sub-picosecond relaxation component. The amplitude exhibits a
sharp drop at the second order phase transition to the 3D ordered CDW state at
T$_{c}^{3D}=183$ K. Above 183 K the amplitude gradually decreases to
\symbol{126}250\ K, and remains constant above that temperature. The behavior
above T$_{c}^{3D}$ is a manifestation of a fluctuating presence of short range
segments with 3D order. This is schematically presented in insets shown by the
charge density modulation along the chains at $T<T_{c}^{3D}$ (long range 3D
order) and at $T>T_{c}^{3D}$ (short range 3D fluctuations when $\xi_{\bot}$
exceeds the interchain separation). b) The temperature dependence of the CDW
state recovery dynamics. Dashed lines present the fits using Eqs.(\ref{NphBCS}%
,\ref{e25}) respectively.}%
\label{BBAmpsFig}%
\end{center}
\end{figure}

Let us now turn to the transient (sub-picosecond) reflectivity
signal, attributed to the photoinduced absorption from the
photoexcited quasiparticles. The T-dependence of the amplitude of
the reflectivity change and the rel. time, determined by the
single exponential decay fit to the
data, is presented in Fig.\ref{BBAmpsFig}.%

\vspace{-0.5cm}

The amplitude of the PI transient has a pronounced temperature
dependence shown in Figure \ref{BBAmpsFig}a). Upon increasing the
temperature, it first slightly increases, followed by a sharp drop
at the second order phase transition to the 3D ordered CDW state
at T$_{c}^{3D}$. Above 183 K the amplitude gradually decreases to
$\sim250$\ K, and remains constant above that temperature all the
way up to the highest temperatures measured ($\sim400$ K). On the
other hand, relaxation time $\tau_{s}$ is roughly constant far
below T$_{c}^{3D}$, shows a quasi-divergence when T$_{c}^{3D}$ is
approached from below, and then drops to constant value of $\sim
0.2$ ps.

The $T$-dependence of the photoinduced signal amplitude below $T_{c}$ has been
fitted with Eq.(\ref{NphBCS}). We have used a BCS functional form for the
T-dependence of the gap consistent with the T-dependence of lattice
distortions \cite{GrunerReview}. Plotting Eq.(\ref{NphBCS}) as a function of
temperature in Fig. \ref{BBAmpsFig}a), we find that the amplitude obtained
from the fits to the data agrees remarkably well with the model for $T<T_{c}$:
amplitude is nearly constant up to nearly 100 K, then increases slightly and
then drops very rapidly near $T_{c}^{3D}$. Using the\ value of the
dimensionless constant $\frac{2\nu}{N(0)\hbar\Omega_{c}}=$ 10 (taking
$N(0)\sim2-3$ eV$^{-1}$spin$^{-1}$cell$^{-1}$ \cite{GrunerBook}, $\Omega
_{c}\sim0.1eV$ \cite{Requardt}, and $\nu\sim$ 1-5, since the gap magnitude is
of the order of 0.1 eV \cite{GrunerReview}), we obtained the value of the gap
$\Delta(0)=850\pm100$ K from the fit of Eq.(\ref{NphBCS}) - in good agreement
with other measurements \cite{GrunerReview,GrunerBook}.

In contrast to the response of the collective modes, both of which disappear
within 10-20 K below $T_{c}$, the amplitude of the quasiparticle transient
gradually drops up to nearly 250 K, and then remains constant up to 400 K. The
polarization anisotropy of the signal above $T>T_{c}^{3D}$ is the same as for
$T<T_{c}$ \cite{TRbb}, strongly suggesting that the origin of the signal above
$T_{c}$ is the same as below $T_{c}$ i.e. photoexcited quasiparticles. The
strong T-dependence of the amplitude above $T_{c}^{3D}$ is inconsistent with
the simple electron-phonon thermalization scenario that was found to explain
the picosecond dynamics in metals \cite{Allen,HF}. In fact, the peculiar
T-dependence of the amplitude above $T_{c}^{3D}$ is suggestive of some
suppression in the electronic DOS, i.e. pseudogap, already above $T_{c}^{3D}$.
We attribute the existence of the pseudogap in the excitation spectrum to the
fluctuating presence of short range 3D segments as depicted schematically in
Figure \ref{BBAmpsFig}a). The 1D CDW correlations along the chains build up at
temperatures far above $T_{c}^{3D}$ (at room temperature the correlation
length along the chains $\xi_{\shortparallel}\approx20$ \AA , exceeding the
lattice constant). On the other hand, in the direction perpendicular to chains
the correlation length $\xi_{\bot}$ is still smaller than the interchain
separation, and the fluctuations are decoupled. Upon cooling, the interchain
correlation length increases and at temperature $T^{\ast}$ exceeds the
distance between two adjacent chains $d_{\bot}$. Below this temperature the
fluctuations have a 3D character with fluctuations on neighboring chains
strongly coupled, giving rise to suppression in the DOS at E$_{F}$. Indeed,
x-ray data on K$_{0.3}$MoO$_{3}$ \cite{Girault} seem to follow this scheme,
suggesting (from the analysis of the Bragg pattern) that 3D fluctuations
develop below $\sim200$ K, i.e. at temperatures $\sim20$ K above $T_{c}^{3D}$
\cite{Girault}. Our data support this picture and suggest that this crossover
temperature is even higher, of the order of $250$ K.

The $T$-dependence of the relaxation time $\tau_{s}$ is shown in
Figure \ref{BBAmpsFig}b). As $T\rightarrow T_{c},$ $\tau_{s}$
appears to \textit{diverge} and then drops to $\tau_{s}\sim0.25$
ps above $T_{c}$. Such behavior is in agreement with the expected
$T$-dependence for the case of a mean-field like gap closing at
$T_{c}^{3D}$. The dashed line in Figure \ref{BBAmpsFig}b) shows a
fit to the data using Eq.(\ref{e25}) with a BCS-like $T$-dependent
gap $\Delta_{BCS}(T)$ with $T_{c}^{3D}=183$ K. Importantly, above
$T_{c}^{3D}$ the relaxation time is constant (within experimental
accuracy), showing no gradual changes between $T_{c}^{3D}$ and 250
K as observed in the T-dependence of the transient amplitude - Fig \ref{BBAmpsFig}%
a). In fact the relaxation time is constant all up to 400K, where
all the CDW fluctuations (even 1D) should be absent, and the
dynamics is governed by electron-phonon thermalization. This gives
further support for association of the gradual decrease in
amplitude above $T_{c}^{3D}$ to the fluctuating presence of short
range segments with 3D order. Namely, if the gap was finite above
$T_{c}^{3D}$ and it was gradually decreasing above $T_{c}^{3D}$
this should be evidenced as a gradual increase in the relaxation
time (according to Eq.(\ref{e25}). Instead, the relaxation
dynamics above $T_{c}^{3D}$ is fast, determined by the fastest
channel of relaxation - in this case electron-phonon
thermalization in metallic regions.

\section{Summary}

We have presented a femtosecond time-resolved optical spectroscopy as a new
experimental tool to probe the changes in the low energy electronic density of
states in strongly correlated electron systems. We showed that the
photoexcited carrier relaxation is strongly affected by the presence of a
small energy gap in the density of states \cite{Kabanov}. The gap creates a
relaxation bottleneck increasing the typically 100 femtosecond relaxation time
to picoseconds. By measuring changes in reflectivity or transmission of the
suitably delayed probe optical pulse one probes the time evolution of the
photoexcited carrier density. The amplitude and the relaxation time of the
fast photoinduced transient strongly depends on temperature and the magnitude
of the single particle gap. Therefore, by analyzing the T-dependences of the
amplitude and relaxation time one can determine the magnitude and temperature
dependence of the gap. Furthermore, since the effective shutter speed is on
the order of a picosecond, the technique is particularly useful to probe the
systems with (dynamic) spatial inhomogeneities.

We applied the femtosecond time-resolved spectroscopy to study a
quasi 1D CDW semiconductor K$_{0.3}$MoO$_{3}$. A fast transient
was found, whose amplitude and relaxation time showed anomalies at
$T_{c}^{3D}$, concurrent with the opening of the Peierls gap. The
amplitude of the fast transient shows an abrupt drop at T$_{c}$,
whereas the relaxation time shows divergence at T$_{c}$ in
agreement with the theoretical model \cite{Kabanov}. The amplitude
of the single particle gap found from the fit to the data was
found to be in agreement with the well established value of
$2\Delta\left(  0\right) \sim1400$ K \cite{GrunerBook}. Above
T$_{c}$ the fast signal amplitude drops gradually to a constant at
$\sim250$ K, which was attributed to the fluctuating presence of
short range segments with 3D CDW order. In addition, amplitude
mode reflectivity oscillations were observed in real time, whose
frequency, and damping are in close agreement with frequency-domain measurements \cite{PougetNeutron,WachterRaman}%
. A $T$-dependent overdamped response is also observed and on the basis of the
$T$-dependence of its damping it is attributed to relaxation of the phason mode.


\begin{chapthebibliography}{99}

\bibitem {Kabanov}V.V. Kabanov, J. Demsar, B. Podobnik and D. Mihailovic,
\textit{Phys. Rev. B} \textbf{59}, 1497 (1999).

\bibitem {ACS}D. Mihailovic and J. Demsar, in \textit{Spectrosopy of
Superconducting Materials}, Ed. Eric Faulques, ACS Symposium Series 730; The
American Chemical Society: Washington, D.C., 1999, p. 230.

\bibitem {TRbb}J. Demsar, K. Biljakovic, D. Mihailovic, \textit{Phys. Rev.
Lett. }\textbf{83}, 800 (1999); see also recent work in the high
excitation regime by A.A. Tsvetkov et al., \textit{Acta Physica
Polonica B} \textbf{34}, 387 (2003).

\bibitem {TR2DCDW}J. Demsar, H. Berger, L. Forro, D. Mihailovic, \textit{Phys.
Rev. B} \textbf{66}, 041101 (2002).

\bibitem {HF}J. Demsar \textit{et al.}, \textit{Phys. Rev. Lett. }\textbf{91,}
027401 (2003).

\bibitem {Averitt}R.D. Averitt \textit{et al.}, \textit{Phys. Rev. B
}\textbf{63}, 140502 (2001).

\bibitem {Averitt2}R.D. Averitt \textit{et al.}, \textit{Phys. Rev. Lett.}
\textbf{87}, 017401 (2001).

\bibitem {JDMgB2}J. Demsar \textit{et al.}, \textit{Phys.
Rev. Lett.} \textbf{91}, 267002 (2003); J. Demsar \textit{et al.},
\textit{Int. J. Mod. Phys.} \textbf{17}, 3675 (2003).

\bibitem {Bi2212}P. Gay \textit{et al.}, \textit{J. Low Temp. Phys.}
\textbf{117}, 1025 (1999).

\bibitem {Tl2201}D.C. Smith \textit{et al.}, \textit{Physica C}
\textbf{341-348}, 2219 (2000).

\bibitem {Hg1223}J. Demsar \textit{et al.}, \textit{Phys. Rev. B }\textbf{63},
54519 (2001).

\bibitem {Schneider}M.L. Schneider \textit{et al.}, \textit{Europhys. Lett.
}\textbf{60} 460 (2002).

\bibitem {Segre}G.P. Segre \textit{et al.}, \textit{Phys. Rev. Lett.}
\textbf{88}, 137001 (2002).

\bibitem {Stevens}C.J. Stevens \textit{et al.}, \textit{Phys.Rev.Lett.
}\textbf{78}, 2212 (1997).

\bibitem {Han}S.G. Han, Z.V. Vardeny, O.G. Symko, G. Koren, \textit{Phys. Rev.
Lett.} \textbf{65}, 2708 (1990).

\bibitem {EPL}J. Demsar \textit{et al.},\textit{\ Europhys. Lett. }%
\textbf{45}, 381 (1999).

\bibitem {Y124}D. Dvorsek \textit{et al.}, \textit{Phys. Rev. B} \textbf{66},
020510 (2002).

\bibitem {ODpaper}J. Demsar \textit{et al.}, \textit{Phys. Rev. Lett.
}\textbf{82}, 4918 (1999).

\bibitem {RothwarfTaylor}A. Rothwarf, B.N. Taylor, \textit{Phys. Rev. Lett.
}\textbf{19}, 27 (1967).

\bibitem {Tsuei}C.C. Tsuei and J.R. Kirtley, \textit{Rev. Mod Phys.}
\textbf{72}, 969 (2000) and the references therein.

\bibitem {Aronov}G.M. Eliashberg, \textit{Zh. Exsp. Theor. Fiz. }\textbf{61},
1274 (1971), A.G. Aronov and B.Z. Spivak, \textit{J. Low Temp. Phys.}
\textbf{29}, 149 (1977).

\bibitem {Lifshitz}J.M. Ziman, \textit{Electrons and Phonons} (Oxford
University Press, London 1960), E.M. Lifshitz, L.P. Pitaevskii,
\textit{Physical kinetics}, (Butterworth-Heinemann ,Oxford, 1995).

\bibitem {GrunerReview}G. Gr\"{u}ner, \textit{Rev.Mod.Phys. }\textbf{60}, 1129 (1988).

\bibitem {GrunerBook}G. Gr\"{u}ner, \textit{Density Waves in Solids},
(Addison-Wesley, 1994).

\bibitem {Xray}J. Graham and A.D. Wadsley, \textit{Acta Cryst. }\textbf{20}, 93, (1966).

\bibitem {PougetNeutron}J.P. Pouget et al., \textit{Phys. Rev. }\textbf{B 43,}
8421 (1991).

\bibitem {WachterRaman}G. Travaglini, I. M\"{o}rke, P. Wachter,
\textit{Sol.State.Comm. }\textbf{45}, 289 (1983).

\bibitem {Zeiger}H.J. Zeiger \textit{et al.}, \textit{Phys. Rev. B
}\textbf{45}, 768 (1992).

\bibitem {KimBB}T.W. Kim \textit{et al.}, \textit{Phys. Rev. }\textbf{B 40,}
5372 (1989).

\bibitem {Tutis}E. Tuti\v{s}, S. Bari\v{s}i\'{c}, Phys. Rev.B \textbf{43}, 8431 (1991).

\bibitem {Degiorgi}L. Degiorgi \textit{et al.}, \textit{Phys.Rev.B}
\textbf{44}, 7808 (1991).

\bibitem {Requardt}H. Requardt \textit{et al.}, \textit{J.Phys.: Cond. Matt.
}\textbf{9}, 8639 (1997).

\bibitem {Allen}P.B. Allen, \textit{Phys. Rev. Lett. }\textbf{59}, 1460 (1987).

\bibitem {Girault}S. Girault, A.H. Moudden, J.P. Pouget, \textit{Phys. Rev.
}\textbf{B 39,} 4430 (1989).
\end{chapthebibliography}
\end{document}